\documentclass[aps,pra,twocolumn,superscriptaddress,nofootinbib]{revtex4-1}
\usepackage{epsfig}
\usepackage{amsmath,amssymb,amsfonts}
\usepackage{amsthm}
\usepackage{graphicx}
\newcommand{\beq}{\begin{equation}}
\newcommand{\eeq}{\end{equation}}
\newcommand{\nn}{\nonumber}

\usepackage{braket}
\newtheorem{theorem}{Theorem}

\newtheorem{proposition}[theorem]{Proposition}

\newcommand{\ba}[2]{\begin{align}\label{#1}#2\end{align}}
\usepackage{mathtools}
\DeclarePairedDelimiter{\floor}{\lfloor}{\rfloor}

\begin{document}

\title{Tree-size complexity of multiqubit states}
\author{\surname{L\^e} Huy Nguy\^en}
\affiliation{Centre for Quantum Technologies, National University of Singapore, 3 Science Drive 2, Singapore 117543, Singapore}
\author{Yu \surname{Cai}}
\affiliation{Centre for Quantum Technologies, National University of Singapore, 3 Science Drive 2, Singapore 117543, Singapore}
\author{Xingyao \surname{Wu}}
\affiliation{Centre for Quantum Technologies, National University of Singapore, 3 Science Drive 2, Singapore 117543, Singapore}
\author{Valerio \surname{Scarani}}
\affiliation{Centre for Quantum Technologies, National University of Singapore, 3 Science Drive 2, Singapore 117543, Singapore}
\affiliation{Department of Physics, National University of Singapore, 2 Science Drive 3, Singapore 117542, Singapore}

\begin{abstract}

Complexity is often invoked alongside size and mass as a characteristic of macroscopic quantum objects. In 2004, Aaronson introduced the \textit{tree size} (TS) as a computable measure of complexity and studied its basic properties. In this paper, we improve and expand on those initial results. In particular, we give explicit characterizations of a family of states with superpolynomial complexity $n^{\Omega(\log n)}= \mathrm{TS} =O(\sqrt{n}!)$ in the number of qubits $n$; and we show that any matrix-product state whose tensors are of dimension $D\times D$ has polynomial complexity $\mathrm{TS}=O(n^{\log_2 2D})$.
\end{abstract}

\begin{widetext}
\maketitle
\end{widetext}

\section{Introduction} 

Quantum mechanics is one of the most tested theories, and it has been verified on different kinds of physical systems. However, when quantum formalism is applied to objects observed in our daily life, something seemingly absurd occurs. The Schr{\"o}dinger's cat presents such a paradox, where an object is placed in a superposition of two macroscopically distinct states, ``dead" and ``alive." Various coherent superpositions claimed to be the ``cat state" were implemented in many physical systems such as mechanical resonators~\cite{Grobacher09,Cleland10}, superconducting qubit~\cite{Friedman00} and heavy molecules \cite{Arndt03,Juffman12}. A number of different measures of effective size are proposed to quantify the macroscopicity of these quantum superpositions \cite{Frowis12,Leggett80,Shimizu02,Bjork04,Korsbakken07,Nimmrichter13}. However, the GHZ state, which maximizes most of the size criteria proposed, is simple in the sense that only a small amount of information is required to describe it. Could complexity be a more suitable criterion for a ``Schr{\"o}dinger's cat"? Indeed, one often finds it stated that complexity is likely to play just as important a role as number and size in testing quantum mechanics at the macroscopic scale. 

Complexity may also be relevant in the context of quantum computing. It is well known that simple quantum states such as matrix product states with bounded bond dimension can be simulated efficiently with classical computers~\cite{Vidal03,Vidal04,Hastings07}. Any state that offers an advantage over classical computing must be significantly complex. We will make it clear below what is meant by ``simple" and ``complex" in the language of complexity measures. It is natural to ask whether there is any connection between complexity and the power of quantum computing. 

As a starting point, we search for complex quantum states. It is often argued that a generic state from the Hilbert space is complex in various senses with high probability \cite{complex}. However, to the best of our knowledge no such states have been explicitly written down (the existing examples of complex states such as the subgroup states in Ref.~\cite{Aaronson04} involve random selection from a subset of the Hilbert space). In this paper, we study an explicit class of states whose complexity grows superpolynomially in the number of qubits. We believe that an explicit construction will be very useful for further studies on the topic of complex quantum states. Among the several complexity measures proposed \cite{Aaronson04,Mora05,Mora07,Benatti06,Rogers08,Yuri12}, we focus on the \textit{tree size} of a quantum state, introduced by Aaronson in an attempt to give a more rigorous foundation to the debate on the possibility of large-scale quantum computing versus a hypothetical breakdown of quantum mechanics \cite{Aaronson04}. This measure of complexity is motivated by the work of Raz, who showed that any multilinear formula for the determinant and permanent of a matrix must be superpolynomial in size~\cite{Raz04}. 

The paper is organized as follows. In Sec.~II we discuss the tree size of a quantum state and its lower and upper bounds. Section III is on the study of an explicit family of superpolynomial complex states for qubits called the determinant and permanent states. Next, we consider the tree-size complexity of matrix product states in Sec.~IV and conclude in Sec.~V.

\section{Tree size of a multiqubit state}

\subsection{Definition}

\begin{figure}[b]
\centering
\includegraphics[scale=0.24]{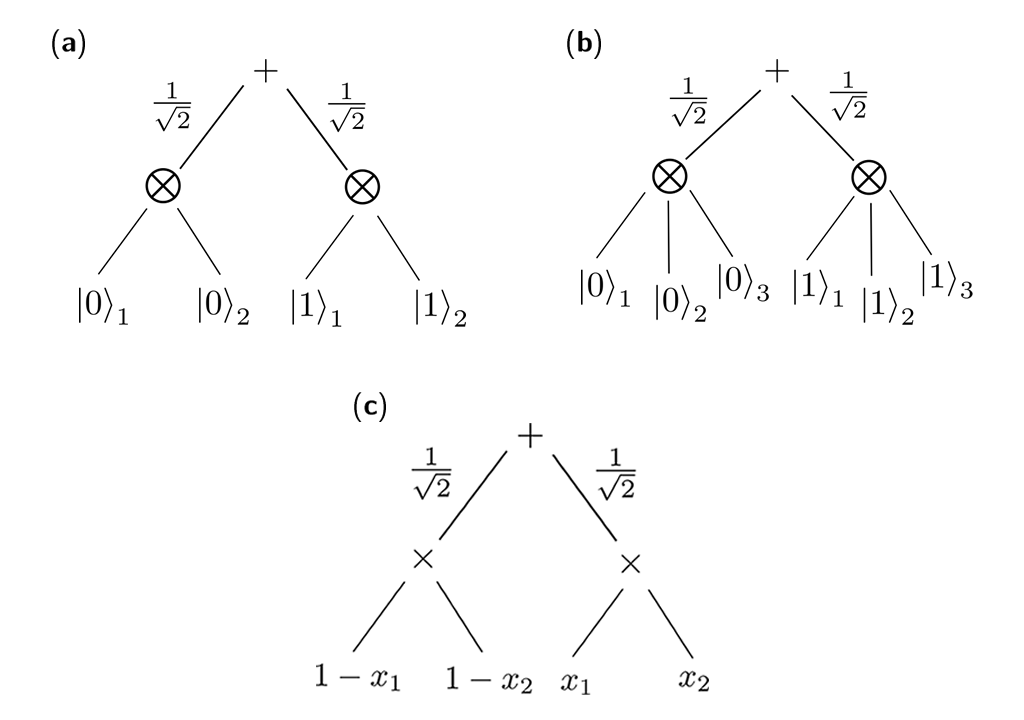}
  \caption{Rooted trees of (a) the Bell state and (b) the three-qubit GHZ state; and (c) a binary tree of the associated multilinear formula that computes the coefficients of the Bell state.}\label{forest}
\end{figure}

Let us first briefly describe the tree size (TS) of a quantum state. An arbitrary pure quantum state of qubits can be represented by a rooted tree (see Fig.~\ref{forest}). Each leaf vertex is labeled with $\alpha \ket{0}+\beta\ket{1}$ for each qubit where $\alpha, \beta$ are complex coefficients; each nonleaf vertex is labeled with either a $+$ gate or a $\otimes$ gate, and complex multiplicative constants are put at the edges of the $+$ gates \cite{Aaronson04}. The rooted trees for the Bell state and the three-qubit GHZ state are given in Fig.~\ref{forest}. The \textit{size} of a rooted tree is defined as the number of leaf vertices. It is obvious that any quantum state can be represented by different rooted trees each with a different size. For example, the  state  $\left(\ket{00}+\ket{01}+\ket{10}+\ket{11}\right)/2$ whose size is 8 can also be written as $\ket{+}\ket{+}$ with size 2. The tree size of a quantum state is taken as the \emph{minimum size over all possible representations}. 

For a quantum state $\ket{\Psi}$, the minimal representation with size $\mathrm{TS}\left(\ket{\Psi}\right)$ is the most compact way of writing down that state. This explains why the tree size is a good measure of complexity: When the most compact form of an object is very complex, it is reasonable to say that the object itself has a high degree of complexity. As an example we consider some of the most commonly encountered $n$-qubit quantum states. It is not difficult to show that the tree size is $O(n)$ for the GHZ state,  $O(n^2)$ for the $W$ state, and $O(n^4)$ for the one-dimensional cluster state \cite{Aaronson04}. The tree size of the two-dimensional cluster state is conjectured to be superpolynomial but no proof has yet been given \cite{Aaronson04}.

We observe that quantum states with polynomial tree size are too simple to be useful for measurement-based quantum computation (MBQC). If a quantum state has a polynomial tree size, its minimal rooted tree can be produced with polynomial computational effort in a classical computer. Moreover, local measurements do not lead to an increase in the tree size of the quantum state~\footnote{After a local projective measurement on a qubit, all the different states at the leaves corresponding to this qubit in the minimal tree are transformed to a single state and hence can be factorized out. It is clear that this operation can only reduce the tree size.}. Therefore, MBQC on a polynomial-tree-size quantum state can be simulated in a classical computer with polynomial overhead and hence offers no significant advantage. This supports the conjecture that the 2D cluster state possesses a superpolynomial tree size. 

At first it may seem from the definition that the tree size depends on the choice of basis $\ket{0}$ and $\ket{1}$, but in fact it is basis independent. This follows immediately from
\begin{proposition}\label{prop1}
If two quantum states can be converted to each other by reversible stochastic local operations and classical communication (SLOCC), then they possess equal tree size. In other words, if there exist invertible local operators (ILO) $A_1,A_2,\ldots,A_n$ such that
\beq
\ket{\psi}=A_1\otimes \dots \otimes A_n \ket{\phi},
\eeq
then $\mathrm{TS}\left(\ket{\psi}\right)=\mathrm{TS}\left(\ket{\phi}\right)$.
\end{proposition}\label{SLOCC}
\begin{proof} Let $\mathrm{mT}_{\phi}$ be the minimal tree of $\ket{\phi}$; applying a local operator $A_i$ on the $i$th qubit simply changes the superposition $\alpha \ket{0}_i + \beta \ket{1}_i$ at an arbitrary leaf of that qubit to a different superposition  $\alpha' \ket{0}_i + \beta' \ket{1}_i$. Thus, the tree $\mathrm{mT}_{\phi}$ can also describe  $\ket{\psi}$, which means $\mathrm{TS}\left(\ket{\psi}\right)\leq\mathrm{TS}\left(\ket{\phi}\right)$. However, we can also write 
\beq
\ket{\phi}=A_1^{-1}\otimes \dots \otimes A_n^{-1} \ket{\psi},
\eeq
and the same line of argument results in $\mathrm{TS}\left(\ket{\phi}\right)\leq\mathrm{TS}\left(\ket{\psi}\right)$. These two inequalities imply that $\mathrm{TS}\left(\ket{\psi}\right)=\mathrm{TS}\left(\ket{\phi}\right)$.\end{proof}

An immediate corollary is that the tree size of a quantum state is basis independent because a change of local basis is equivalent to applying local unitary operators which are a special case of invertible local operators. Another implication of Proposition \ref{prop1} is that \textit{all the states belonging to a SLOCC-equivalent family must have the same tree size}. This result is useful for finding the tree size for quantum states of a few qubits since it is possible to examine all the SLOCC-equivalent families when the number of qubits is small \cite{Dur00,Lamata07}.

\subsection{Upper bounds on the tree size}

While it is not easy to compute the tree size of a given state, nontrivial upper and lower bounds are obtainable. An arbitrary $n$-qubit quantum state can be written as a superposition of the computational basis states
\beq\label{gex}
\ket{\Psi}=\sum_{x=0}^{2^n-1} \alpha_x\ket{x},
\eeq
where each of the terms $\ket{x}$ is an $n$-bit string $\ket{x_1,x_2,...,x_n}$ with $x_i=0$ or $1$. Therefore, its TS is upper bounded by $n2^n$ \cite{Aaronson04}. In fact, it is easy to improve on this bound by collecting all the terms $\ket{x}$ with $x_1=0$ to one group and those with $x_1=1$ to another group; one may write the $n$-qubit state in Eq.~\eqref{gex} as $\ket{\Psi}=\ket{0} \ket{\chi_0}+\ket{1}\ket{\chi_1}$, where $\ket{\chi_0}$ and $\ket{\chi_1}$ are some states of $n-1$ qubits. Let $B_n$ denote the size of an $n$-qubit state when written in the form of the above equation, we have $B_n=2(B_{n-1}+1)$. Solving this recursive formula with $B_1=1$ yields the upper bound
\beq
\mathrm{TS}_n\,\leq\, 3 \times 2^{n-1}-2\,.
\eeq It may be possible to reduce this bound further using an optimized decomposition \cite{Acin00}. For a given state, an upper bound can always be constructed by studying an explicit decomposition, but we do not know yet any way to estimate its tightness.

\subsection{Lower bounds on the tree size}

While upper bounds are sufficient to prove that a state is not complex, \textit{lower bounds} are needed in order to prove that a state is complex. One of the reasons why the tree size is appealing is that rigorous lower bounds can actually be computed.

As mentioned earlier, the tree-size complexity measure for quantum states is closely related to the size of multilinear formulas (MFS). A multilinear formula of the complex variables $\{x_1,x_2,...,x_n\}$ can be represented by a binary tree with each leaf vertex labeled with a variable $x_k$ or a complex constant \cite{Raz04}. The nonleaf vertices are labeled with either $+$ or $\times$. The size of each binary tree is the number of leaf vertices. The size of a multilinear formula is the minimum size taken over all possible binary-tree representations. 

Now we define an associated multilinear formula $f_{\psi}(x_1,x_2,...,x_n)$ that maps each bit string $x=\{x_1,x_2,...,x_n\}$ to the coefficient $\alpha_x$, that is,
\beq
f_{\psi}(x)=\alpha_x.
\eeq
A binary tree for $f_{\psi}$ can be obtained from a rooted-tree of the corresponding state $\ket{\Psi}$ by a straightforward procedure: From the rooted tree of the quantum state, one replaces each $\ket{0}_i$ by $1-x_i$, $\ket{1}_i$ by $x_i$, and the $\otimes$ gates by a set of binary $\times$ gates \cite{Aaronson04}. It can be verified that the resulting multilinear formula indeed computes the coefficients $\alpha_x$ of the state $\ket{\Psi}$.  The example for the Bell state is shown in Fig.~\ref{forest}(c). It is therefore clear that, from the minimal tree of state $\ket{\Psi}$, one can immediately obtain a multilinear formula that computes its coefficients (even though there may exist such a formula with smaller size~\footnote{For instance, take the state $\frac{1}{\sqrt{2}}\left(\ket{00}-\ket{11}\right)$: The direct recipe gives the multilinear formula $\frac{1}{\sqrt{2}}\left[(1-x_1)(1-x_2)-x_1x_2\right]$, which has size 4; but this can be further expanded to $\frac{1}{\sqrt{2}}\left(1-x_1-x_2\right)$, which has size 2. Also, we notice that in Ref.~\cite{Raz04} the size of a binary tree is defined as the number of nodes $N$, which is related to the number of leaves $L$ by $N=2L-1$.}). This implies that $\mathrm{MFS}(f_{\psi})=O\left[\mathrm{TS}\left(\ket{\Psi}\right)\right]$: In particular, \textit{if $f(x)$ has a superpolynomial $\mathrm{MFS}$, then the state $\sum_x f(x)\ket{x}$ has a superpolynomial tree size}. This is one of the main results stated in Theorem 4 of Ref.~\cite{Aaronson04}.

\section{Superpolynomial complex quantum states} 

In this section we write down explicitly a class of states that possesses superpolynomial tree size and discuss their properties. These states are immediate consequences of Theorem 4 in Ref.~\cite{Aaronson04} and the superpolynomial lower bound proved in Ref.~\cite{Raz04}. 

\subsection{The family of states}

Raz showed in Ref.~\cite{Raz04} that any multilinear formula that computes the determinant or the permanent (or in fact any immanant with nonzero coefficients \footnote{An immanant with nonzero coefficients $c(\sigma)$ of an $m\times m$ matrix $M$ is $\mathrm{Imm}\{M\}=\sum_{\sigma}c(\sigma) \prod_{i=1}^m M_{i\sigma_i}$ \cite{Littlewood50}. Raz does not study this case explicitly; however, the argument in Sec.~7 of Ref.~\cite{Raz04} can be used to show that the partial-derivatives matrix $M_{\phi_A}$ corresponding to a nonzero-coefficient immanant has full rank. Hence, the multilinear formula size of these immanants is also $m^{\Omega(\log m)}$.}) of a $m\times m$ matrix must have superpolynomial tree size $m^{\Omega(\log m)}$. 

Based on this result, we can exhibit an explicit family of superpolynomial complex multiqubit states when the number of qubits is $n=m^2$ with $m$ a positive integer. Since the construction is analogous for the determinant, the permanent, and any immanant, we focus on the \textit{determinant state} unless otherwise specified.

For the construction, the qubits are first labeled as $x_{11},x_{12},...,x_{mm}$ and arranged to a matrix
\begin{eqnarray}
\{x\} = 
\begin{pmatrix}
x_{11}     &x_{12} & \cdots & x_{1m} \\
x_{21} &x_{22}    & \cdots & x_{2m} \\
\vdots  &\vdots    & \ddots & \vdots \\
x_{m1}        &x_{m2}    &  \cdots      & x_{mm}
\end{pmatrix}.
\end{eqnarray}
So, each bit string $\ket{x}$ in the expansion of Eq.~\eqref{gex} is associated with a (0,1) matrix $\{x\}$ whose elements are $0$ or $1$.  We call the state $\sum_{x} \alpha_x\ket{x}$ the determinant state if the coefficient $\alpha_x$ is taken as the determinant of the corresponding (0,1) matrix. In other words, the $m^2$-qubit determinant state is
\ba{nex}{
\ket{\mathrm{det}_m}&=\sum_{x=0}^{2^n-1}\mathrm{det}(\{x\})\ket{x},}
where the normalization constants are neglected. The determinant states for $m=1,2$ are
\ba{}{
\ket{\mathrm{det}_1}=&\ket{1}, \nn \\
\ket{\mathrm{det}_2}=&-\ket{0110}-\ket{0111}+\ket{1001}\nn \\
&+\ket{1101}+\ket{1101}-\ket{1110}.}
Since $\mathrm{MFS}(f_{\psi})=O\left[\mathrm{TS}\left(\ket{\Psi}\right)\right]$, the tree size of $\ket{\mathrm{det}_m}$ must scale as $m^{\Omega(\log m)}=n^{(1/4)\Omega(\log n)}$, which is superpolynomial in $n$.

\subsection{Upper bound for the tree size}

The size of the representation of $\ket{\mathrm{det}_m}$ given in Eq.~\eqref{nex} is enormous for large $m$. There are $2^{m^2}$ different $(0,1)$ $m\times m$ matrices. At first sight it may seem that the determinant of $(0,1)$ matrices must be small, yet the largest possible determinant is $2^{-m}\sqrt{(m+1)^{m+1}}$ with equality if and only if there is a Hadamard matrix of order $m+1$ \cite{Williamson45,Zivkovic06}. The Hadamard conjecture states that this is true for $m=3 \, (\mathrm{mod}\, 4)$ but the search for a proof remains a long-standing unsolved problem in mathematics \cite{Brenner72}. Moreover, the number of terms left in the expansions of Eq.~\eqref{nex} is equal to the number of $m\times m$ (0,1) matrices with non vanishing determinant, which is also unknown for large $m$ \cite{Metropolis67}. For $m\leq 8$ numerical computation shows that this number is larger than $0.3 \times 2^{m^2}$ and hence it grows rapidly with increasing $m$ \cite{Zivkovic06}. 

A much more compact representation than Eq.~\eqref{nex} is obtained by using the expansion by minors (Laplace expansion) \cite{Meyer01}
\beq
\det{\{x\}}=\sum_{j=1}^{m}(-1)^{i+j}x_{ij}M_{ij} ,
\eeq
where $M_{ij}$ is the determinant of $\{x\}$ with row $i$ and column $j$ crossed out. Inserting this expression for the determinant into Eq.~\eqref{nex} and summing over all possible values of $x$ yields the recursive formula
\ba{rfm}{\ket{\mathrm{det}_m}&=\sum_{j=1}^{m}(-1)^{i+j}\ket{1}_{ij}\ket{\mathrm{det}_{m-1}}_{ij}\ket{+}^{\bigotimes (2m-2)},}
which is understood as follows: The qubit at row $i$ and column $j$ is in the $\ket{1}$ state, the qubits in the $(m-1)\times (m-1)$ block obtained by removing row $i$ and column $j$ is in the $\ket{\mathrm{det}_{m-1}}$ state, and the remaining $2m-2$ qubits are in the $\ket{+}$ states. Starting from $\ket{\det_{1}}=\ket{1}$, one may use the recursive formula to generate determinant states with larger numbers of qubits, for instance,
\beq
\ket{\mathrm{det}_2}=\ket{1}\ket{+}\ket{+}\ket{1}-\ket{+}\ket{1}\ket{1}\ket{+}.
\eeq
The permanent state also adopts the recursive formula in Eq.~\eqref{rfm} with all the $-1$ signs switched to $+1$.

How compact is this new expression? Let us denote by $S_m$ the size of $\ket{\det_m}$ when written as in Eq.~\eqref{rfm}. By counting the number of leaves we see that
\ba{}{
S_m=m(S_{m-1}+2m-1),}
from which we obtain $S_m=m!\lambda_m$ with $\lambda_m=\sum_{k=1}^{m}\frac{2k-1}{(k-1)!}$ which converges rapidly to $3e$. Thus, the tree size of $\ket{\det_m}$ is
\beq\label{upperbound}
\mathrm{TS}(\ket{\mathrm{det}_m})=O\left(m!\right).
\eeq
This is the smallest upper bound we were able to find, which suggests the possibility that the lower bound $m^{\Omega(\log m)}$ can be improved further.

\subsection{Entanglement properties}

A less compact but more useful expression for studying the entanglement properties of the determinant state is based on the Leibnitz formula \cite{Meyer01}
\ba{}{
\det{\{x\}}=\sum_{\sigma}\mathrm{sgn}(\sigma) \prod_{i=1}^m x_{i\sigma_i},}
where the summation is taken over all $m!$ possible permutation $\sigma$ of the set $\{1,2,...,m\}$; the sign is $-1$ for odd and $+1$ for even permutations. With the help of this formula Eq.~\eqref{nex} can be rewritten as
\ba{psex}{
\ket{\mathrm{det}_m}=\sum_{\sigma}\mathrm{sgn} (\sigma)\left(\bigotimes_{i=1}^{m}\ket{1}_{i\sigma_i}\right)\ket{+}^{\bigotimes (m^2-m)}.} Again, the permanent state has a similar representation with the only difference being the omission of the $\mathrm{sgn}(\sigma)$ terms.

The above formula expresses the determinant state as a sum of $m!$ product states. It is also a minimal product-state expansion \cite{Eisert01}. The reason is that any two terms in this expansion have at least a pair of qubits that is in the state $\ket{1}\ket{+}$ in one term and $\ket{+}\ket{1}$ in the other; and any combination of $\ket{1}\ket{+}$ and $\ket{+}\ket{1}$ is a mixed state with respect to either qubit in the pair. Thus, it is not possible to have a product-state expansion that has less than $m!$ terms \cite{Briegel01}. Thence, the Schmidt measure of the determinant state is $\log_2 (m!)$. This measure is a useful tool for quantifying multiparty entanglement \cite{Hein04}. We see that if the Schmidt measure of an $n$-qubit quantum state is $E_S$, then the minimal product-state expansion has $2^{E_S}$ terms and hence the tree size of this state is bounded above by $n2^{E_S}$. As a consequence, a very complex quantum state (with large tree size) must also be highly entangled. The Schmidt measure is also a lower bound of the entanglement persistency which is defined as the smallest number of local measurements needed to disentangle a quantum state with certainty \cite{Briegel01,Eisert01}. An examination of local measurements with the help of the recursive formula in Eq.~\eqref{rfm} strongly suggests that the entanglement persistency of the determinant state is $n-1$ which is the maximal value achievable for an $n$-qubit quantum state.  

\subsection{Some symmetry properties}

The determinant and permanent states are highly symmetric. The determinant of a matrix changes only its sign under the interchange of any two rows or any two columns while the permanent remains unchanged. If one defines $R_{ij}$($C_{ij}$), $i\neq j$, as the unitary operation that swaps the quantum state of the $m$ qubits in row (column) $i$ with that of the $m$ qubits in row (column) $j$, it can be verified that
\ba{}{
R_{ij}\ket{\mathrm{det}_m}&=C_{ij}\ket{\mathrm{det}_m}=-\ket{\mathrm{det}_m},\nn \\
R_{ij}\ket{\mathrm{per}_m}&=C_{ij}\ket{\mathrm{per}_m}=\ket{\mathrm{per}_m},}
so the determinant (permanent) state is the eigenvector with eigenvalue $-1$ (1) of the operators $R_{ij}$ and $C_{ij}$. One may also check that both the determinant and permanent states are invariant under the matrix transposition operation.

\subsection{A quantum circuit generating complex states} 

Before considering a quantum circuit that generates states with superpolynomial tree size, it is worth mentioning that the subgroup states, which are proved to be superpolynomial in tree size, can be generated by a polynomial time circuits \cite{Aaronson04}. Therefore, it is a possibility that a superpolynomial complex state in the sense of tree size can be realized by a simple quantum circuit.

\begin{figure}[t]
\centering
\includegraphics[scale=0.3]{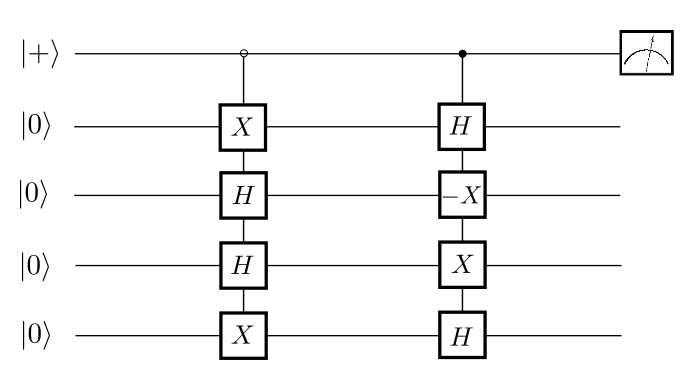}
  \caption{A circuit with controlled operations and ancilla qubits for creating the $four$-qubit determinant state, or a state which is equivalently complex.}\label{circuit}
\end{figure}

Any state with a product-state expansion like that in Eq.~\eqref{psex} can be realized with the help of ancilla qubits \cite{Mora05}. Let $s$ be the smallest integer such that $2^s \geq m!$. We first prepare $s$ ancilla qubits in the equal superposition $\ket{+}_a^{\otimes s}$ and $m^2$ main qubits in the initial product state $\ket{\Phi_0}=\ket{0}^{\otimes m^2}$. The total state is $\sum_{i=1}^{N} \ket{i}_a \ket{\Phi_0}$ where $N=2^s$. Multi controlled Hadamard and Pauli gates are then used to create the state 
\beq
\sum_{i=1}^{m!} \ket{i}_a \ket{\Phi_i} +\sum_{i=m!+1}^{N}\ket{i}_a \ket{\Phi_0},
\eeq
where $\ket{\Phi_i}$ is one of the product states in the expansion of Eq.~\eqref{psex}. Note that for the last $N-m!$ states $\ket{i}_a$ of the ancilla qubits, we do nothing to the main qubits. Next, the ancilla qubits are measured in the $\{+,-\}$ basis. For any outcome one may verify that the resulting state is 
\beq
\ket{\Psi_m}=\sum_{i=1}^{m!}c(i)\ket{\Phi_i}+\left(\sum_{i=m!+1}^{N}c(i)\right)\ket{\Phi_0},
\eeq 
where the $c(i)$ coefficients are equal to either $+1$ or $-1$. How these signs distribute among the terms depends on the specific measurement outcome. The state described by the first summation in the above equation can be rewritten after the change of index $i\rightarrow \sigma$ as
\beq\label{immanant}
\ket{\Psi_1}=\sum_{\sigma}c(\sigma)\left(\bigotimes_{i=1}^{m}\ket{1}_{i\sigma_i}\right)\ket{+}^{\bigotimes (m^2-m)},
\eeq
whose associated multilinear formula is the matrix immanant with nonzero coefficients
\beq
\mathrm{Imm}\{x\}=\sum_{\sigma}c(\sigma) \prod_{i=1}^m x_{i\sigma_i}.
\eeq
Since the quantum state $\ket{\Psi_m}$ is the sum of $\ket{\Psi_1}$ with tree size $m^{\Omega(\log m)}$ and a product state with tree size $m^2$, it follows that the tree size of this state must also be $m^{\Omega(\log m)}$ \footnote{If $\mathrm{TS}(\Psi_m) \leq m^{\epsilon \log m}$ for some positive constant $\epsilon$,  then $\mathrm{TS}(\Psi_1) \leq m^{\epsilon \log m} + m^2$. Thus, for any constant $\mu>\epsilon$ we have $\mathrm{TS}(\Psi_1) < m^{\mu \log m}$ for sufficiently large $m$, which is a contradiction.}. In short, we obtain a superpolynomial complex quantum state regardless of the measurement outcomes. It is obvious that the determinant and permanent states are two special examples of the immanant states given in Eq.~\eqref{immanant}. The quantum circuit for the $m=2$ case is shown in Fig.~\ref{circuit}. 

\subsection{A frequent misunderstanding}

We finish this section by addressing explicitly a misunderstanding that we have frequently encountered when presenting this work. It is a well-known fact that optimal algorithms make the determinant much easier to compute than the permanent. Based on this observation, our study, in which the determinant and the permanent states are treated on equal footing, may look suspicious. In reality, a comparison cannot be drawn. Concretely:

(1) The difference in computability does not cast any shadow on Raz's proof that both the determinant and the permanent require a multilinear formula of similar, superpolynomial complexity: It just means that the optimal algorithms do not use multilinear formulas. Given Raz's bounds, the bound on the complexity of the states follows rather immediately, as explained above.

(2) In order to prepare (say) the permanent state, one does not need to be able to compute permanents. Also, the permanent state is not a resource that would allow a simple computation of the permanent: Indeed, in order to read the value of the permanent from the corresponding state, one needs to estimate a probability that is exponentially small in the size of the matrix. In other words, it may be possible to find a simple circuit that produces those states, without any relation with the complexity of computing the formula, and without trivializing computationally complex problems. This is the same situation as in ``boson sampling" \cite{bosonsampling}. There, the output states have coefficients which are proportional to the permanent of matrices; nevertheless, the circuit that prepares those states can be designed without any knowledge of the value of the permanents, and the sampling does not lead to an efficient computation of the permanent.

\section{Tree size of matrix product states}

Matrix product states have attracted a lot of interest due to their applications in quantum information theory and ground-state computation in one-dimensional systems \cite{Perez07}. In this section we discuss a relation between the bond dimension of a matrix product state and its tree size. It turns out that matrix product states with bounded bond dimension are ``simple" as their tree size is polynomial in the number of qubits. In other words, when it comes to storing quantum states, matrix product representation with bounded bond dimension does not offer a superpolynomial advantage over the usual bra-ket notation.   

\begin{proposition}
For a quantum state $\ket{\psi_n}$ written in matrix product representation (MPS) with open boundary condition (OBC) \cite{Perez07} as
\beq\label{MPS}
\ket{\psi_n}=\bigotimes_{i=1}^{n}(A_0^{(i)}\ket{0}+A_1^{(i)}\ket{1}),
\eeq
where $A_i^{(1)}$ are $1\times D$ matrices (row vectors), $A_i^{(n)}$ are $D\times 1$ matrices (column vectors), and the rest are $D\times D$ matrices, the tree size is $O\left[n^{\log_2 (2D)}\right]$. 
\end{proposition}
The tensor product in Eq.~\eqref{MPS} must be written in the correct order from $1$ to $n$ due to non-commutativity. It is obvious from this proposition that the tree size is polynomial when the bond dimension $D$ is bounded.

\begin{proof} By writing the $D\times D$ identity matrix as 
\beq
I=\sum_{s=1}^D e_s e_s^{T},
\eeq
where $e_s$ is the $s$th unit column vector, and inserting it between the $n/2$ and $n/2+1$ terms in the product of Eq.~\eqref{MPS} we arrive at
\beq
\ket{\psi_n}=\sum_{s=1}^D \ket{\phi^{s,1}_{n/2}} \ket{\phi^{s,2}_{n/2}},
\eeq
where
\ba{}{
\ket{\phi^{s,1}_{n/2}}&=\bigotimes_{i=1}^{n/2}(A_0^{(i)}\ket{0}+A_1^{(i)}\ket{1})e_s,\nn \\
\ket{\phi^{s,2}_{n/2}}&=e_s^{T}\bigotimes_{i=\frac{n}{2}+1}^{n}(A_0^{(i)}\ket{0}+A_1^{(i)}\ket{1})}
are $n/2$ qubit quantum states. Therefore,
\beq 
\mathrm{TS}\left(\ket{\psi_n}\right)~\leq~2D \times  \mathrm{TS}\left(\ket{\phi^{s,k}_{n/2}}\right)
\eeq
and this recursive relation yields $\mathrm{TS}\left(\ket{\psi_n}\right)=O\left[n^{\log_2 (2D)}\right]$. \end{proof}
 
A direct corollary of this is that the 1D cluster state, which has an OBC-MPS representation with $D=2$ \cite{Perez07}, can be described by a tree with $O(n^2)$ leaves. This is an improvement over the existing upper bound $O(n^4)$ given in Ref.~\cite{Aaronson04}.
 
In the above we have assumed that the number of qubit is a power of two. When this is not true, one simply inserts the identity matrix between the first $2^{\floor{\log_2 (n)}}$ qubits and the rest. Repeating the process yields
\beq 
\mathrm{TS}\left(\ket{\psi_n}\right)\leq(2D)^{\floor{\log_2(n)}+1}.
\eeq
Therefore, the asymptotic upper bound $O\left[n^{\log_2 (2D)}\right]$ still holds. Finally, in the more general case when the matrices $A^{(i)}$ have dimension $D_i \times D_{i+1}$ that may differ from qubit to qubit, the same line of argument also results in $\mathrm{TS}\left(\ket{\psi_n}\right)=O\left[n^{\log_2 (2D)}\right]$ with $D=\max{\left(D_i\right)}$. 

\section{Conclusion} To conclude, we were able to construct explicitly a class of superpolynomial complex $n$-qubit quantum states when $n$ is a square number. The tree size of these states, which we call the determinant and permanent states, has a lower bound of $n^{\Omega(\log n)}$. The best upper bound we are able to find for the tree size is $O(\sqrt{n}!)$ as given in Eq.~\eqref{upperbound}. These states are special cases of a wider family of superpolynomial-tree-size quantum states which are constructed based on the immanant of a matrix. These states are highly symmetric due to the invariance of the determinant and permanent with respect to the interchange of any two rows or two columns. A quantum circuit for realizing a quantum state with superpolynomial tree size is proposed. We also prove that all states in a SLOCC-equivalent class has the same tree size, and the tree size of matrix product states with bounded bond dimension is polynomial. 

\begin{acknowledgements} We are grateful to S. Aaronson, I. Arad, D. Cavalcanti, W. D\"ur, and T. Lee for helpful comments and stimulating discussions. This work is supported by the Centre for Quantum Technologies (CQT). CQT is a Research Centre of Excellence funded by Ministry of Education and National Research Foundation of Singapore.
\end{acknowledgements}


\begin{thebibliography}{99}

\bibitem{Grobacher09}
S. Gr{\"o}blacher, K. Hammerer, M.R. Vanner and M. Aspelmeyer, Nature (London) \textbf{460}, 724 (2009).

\bibitem{Cleland10}
A.D. O'Connell, M. Hofheinz, M. Ansmann, R.C. Bialczak, M. Lenander, E. Lucero, M. Neeley, D. Sank, H. Wang, M. Weides, J. Wenner, J.M. Martinis, and A.N. Cleland, Nature (London) \textbf{464}, 697 (2010).

\bibitem{Friedman00}
J. Friedman, V. Patel, W. Chen, S. Tolpygo and J. Lukens, Nature (London) \textbf{406}, 43 (2000).

\bibitem{Arndt03}
L. Hackerm\"uller, S. Uttenthaler, K. Hornberger, E. Reiger, B. Brezger, A. Zeilinger, and M. Arndt, Phys. Rev. Lett. \textbf{91}, 090408 (2003).

\bibitem{Juffman12}
T. Juffmann, A. Milic, M. M{\"u}llneritsch, P. Asenbaum A. Tsukernik, J. T{\"u}xen, M. Mayor, O. Cheshnovsky, and M. Arndt, Nat. Nanotechnol. \textbf{7}, 297 (2012).

\bibitem{Frowis12}
F. Fr\"owis and W. D\"ur, New J. Phys. \textbf{14}, 093039 (2012).

\bibitem{Leggett80}
A.J. Leggett, Prog. Theor. Phys. Suppl. \textbf{69}, 80 (1980).

\bibitem{Shimizu02}
A. Shimizu and T. Miyadera, Phys. Rev. Lett. \textbf{89}, 270403 (2002).

\bibitem{Bjork04}
G. Bi\"ork G and P.G.L Mana, J. Opt. B: Quantum Semiclass. Opt. textbf{6}, 429 (2004).

\bibitem{Korsbakken07}
J.I. Korsbakken, K.B. Whaley, J. Dubois and J.I. Cirac, Phys. Rev. A \textbf{75}, 042106 (2007).

\bibitem{Nimmrichter13}
S. Nimmrichter, K. Hornberger, Phys. Rev. Lett \textbf{110}, 160403 (2013).

\bibitem{Vidal03}
G. Vidal, Phys. Rev. Lett. \textbf{91}, 147902 (2003).

\bibitem{Vidal04}
G. Vidal, Phys. Rev. Lett. \textbf{93}, 040502 (2004).

\bibitem{Hastings07}
M.B. Hastings, J. Stat. Mech. (2007), P08024.

\bibitem{complex}
The vast majority of many-body quantum states in the Hilbert space can only be produced after an exponentially long time by local Hamiltonians [D. Poulin, A. Qarry, R. Somma, and F. Verstraete, Phys. Rev. Lett. \textbf{106}, 170501 (2011)], or have near-maximal entanglement [P. Hayden, D.W. Leung, and A. Winter, Comm. Math. Phys. \textbf{265}, 95 (2006); D. Gross, S.T. Flammia, and J. Eisert, Phys. Rev. Lett. \textbf{102}, 190501 (2009)].


\bibitem{Aaronson04}
S. Aaronson, STOC '04 Proceedings of the 36th Annual ACM Symposium on Theory of Computing, (ACM, New York, 2004), pp. 118-127; arXiv:quant-ph/0311039.

\bibitem{Dur00}
W. D\"ur, G. Vidal, and J. I. Cirac, Phys. Rev. A \textbf{62}, 062314 (2000).

\bibitem{Lamata07}
L. Lamata, J. Le\'on, D. Salgado, and E. Solano, Phys. Rev. A \textbf{75}, 022318 (2007).
 
\bibitem{Mora05}
C.E. Mora and H.J. Briegel, Phys. Rev. Lett. \textbf{95}, 200503 (2005).

\bibitem{Mora07}
C. E. Mora, H. J. Briegel and B. Kraus, Int. J. Quantum Inform. \textbf{5},  729 (2007).

\bibitem{Benatti06}
F. Benatti, Nat. Comput. \textbf{6}, 133 (2006).

\bibitem{Rogers08}
C. Rogers, V. Vedral and R. Nagarajan, Int. J. Quantum Inform. \textbf{6}, 907 (2008).

\bibitem{Yuri12}
Y.C. Campbell-Borges and J.R.C. Piqueira, Int. J. Quant. Inf. \textbf{10}, 1250047 (2012).

\bibitem{Raz04}
 R. Raz, Proc. ACM STOC, 633 (2004).
 

\bibitem{Acin00}
A. Ac\'{i}n,  A. Andrianov, L. Costa, E. Jan\'{e}, J.I. Latorre, and R. Tarrach, Phys. Rev. Lett. \textbf{85}, 1560 (2000).

\bibitem{Littlewood50}
D.E. Littlewood, The Theory of Group Characters and Matrix Representations of Groups (Oxford University Press, New York, 1950), p. 81.

\bibitem{Williamson45}
 J. Williamson, Amer. Math. Monthly \textbf{52}, 417 (1945).
 
\bibitem{Zivkovic06}
M. \v{Z}ivkovi\'{c}, Linear Algebra Appl. \textbf{414}, 310 (2006).

\bibitem{Brenner72}
J. Brenner, Amer. Math. Monthly \textbf{79}, 626 (1972).

\bibitem{Metropolis67}
N. Metropolis and P.R. Stein, J. Comb. Theory \textbf{3}, 191 (1967).

\bibitem{Meyer01}
C.D. Meyer, \emph{Matrix Analysis and Applied Linear Algebra} (SIAM, 2001), p. 549.

\bibitem{Eisert01}
J. Eisert and H.J. Briegel , Phys. Rev. A \textbf{64}, 022306 (2001).

\bibitem{Briegel01}
H.J. Briegel and R. Raussendorf,  Phys. Rev. Lett. \textbf{86}, 910 (2001).

\bibitem{Hein04}
M. Hein, J. Eisert, and H.J. Briegel, Phys. Rev. A. \textbf{69}, 062311 (2004).



\bibitem{bosonsampling} S. Aaronson and A. Arkhipov, Proceedings of the 43rd Annual ACM Symposium on Theory of Computing, (ACM, New York, 2011), p. 333; M.A. Broome \emph{et. al}, Science \textbf{339}, 794 (2012). 

\bibitem{Perez07}
D. Perez-Garcia, F. Verstraete, M.M. Wolf, and J.I. Cirac, Quantum Inf. Comput. \textbf{7}, 401 (2007).



\end{thebibliography}
\end{document}